\documentclass[12pt,doublespace]{article}
\usepackage{hyperref}
\usepackage{amsmath}
\usepackage{graphicx}
\usepackage{color}
\usepackage{amssymb} 

\definecolor{myblue}{rgb}{0.14,0.11,0.49}
\definecolor{myred}{rgb}{0.74,0.22,0.15}
\definecolor{mygreen}{rgb}{0.05,0.52,0.42}
\definecolor{myyellow}{rgb}{0.96,0.92,0.13}
\definecolor{myorange}{rgb}{1,0.61,0.36}
\definecolor{mypurple}{rgb}{0.71,0.02,1}

\definecolor{htc}{rgb}{1,1,1} 

\newcommand{\Couleur}[1]{\textcolor{myblue}{#1}}

\def\be{\begin{equation}}
\def\ee{\end{equation}}
\def\bea{\begin{eqnarray}}
\def\eea{\end{eqnarray}}
\def\dd{\mathrm{d}}

\date{}

\title{Quantum mechanics for three versions of the Dirac equation in a curved spacetime}
\author{Mayeul Arminjon$^1$ and Frank Reifler$^2$\\
$^1$ \small\it CNRS (Section of Theoretical Physics).\\
\small\it At Laboratory ``Soils, Solids, Structures---Risks,"\\ \small\it CNRS \& University of Grenoble, Grenoble, France.\\ \\
$^2$ \small\it Lockheed Martin Corporation, \\
\small\it Maritime Systems \& Sensors,\\ 
\small\it Moorestown, New Jersey, USA.}

\begin{document}

\maketitle          


  \section{Quantum Mechanics in a gravitational field}
\vspace{3mm}
\begin{itemize}
	
{
		\item Quantum effects in the classical gravitational field {\it are observed,} e.g. in neutron interferometry \cite{COW1975,WernerStaudenmannColella1979}. All of the effects observed until now had been previously predicted by using the non-relativistic Schr\"odinger equation in the Newtonian gravity potential \cite{WernerStaudenmannColella1979,OverhauserColella1974,LuschikovFrank1978}, and are still described by this same non-relativistic approximation.\\
		
		\item However, gravity is currently described by relativistic theories with {\it curved spacetime}.\\
		
		\item The usual way to write the wave equations of quantum mechanics (QM) in a curved spacetime is by {\it covariantization} (this is connected with the {\it equivalence principle}): at any given event $\Couleur{X}$, the sought-for equation in a curved spacetime should coincide with the flat-spacetime version in coordinates where the connection vanishes at $\Couleur{X}$ and $\Couleur{g_{\mu \nu }(X)=\eta _{\mu \nu }}$. \\
		
		\item For the Dirac equation with standard (spinor) transformation, this leads to the {\it standard equation (Dirac-Fock-Weyl or DFW)}, which does not obey the equivalence principle, at least not the {\it genuine} one \cite{A39}.\\
		
\item In a previous work, {\it two alternative equations} were got by {\it applying directly the classical-quantum correspondence} \cite{A39}. Thus, we have indeed three different a priori inequivalent versions of the Dirac equation in a curved spacetime.\\
} 
\end{itemize}

  \section{Three Dirac equations in a curved spacetime}
{
The three gravitational Dirac equations have the same {\it form:}
\be\label{Dirac-general}
\Couleur{\gamma ^\mu D_\mu\psi=-im\psi},\\
\ee
\noindent where \Couleur{$\gamma ^\mu =\gamma ^\mu (X)$} (\Couleur{$\mu =0,...,3$}) is a field of \Couleur{$4\times 4$} complex matrices defined on the spacetime \Couleur{V} [endowed with a Lorentzian metric $\Couleur{g_{\mu \nu}}$, with inverse matrix $\Couleur{(g^{\mu \nu})}$], such that 
\be \label{Clifford}
\Couleur{\gamma ^\mu \gamma ^\nu + \gamma ^\nu \gamma ^\mu = 2g^{\mu \nu}\,{\bf 1}_4, \quad \mu ,\nu \in \{0,...,3\}} \quad (\Couleur{{\bf 1}_4\equiv \mathrm{diag}(1,1,1,1)});
\ee
and where \Couleur{$\psi$} is a {\it bispinor} field for the standard, {\it DFW} equation, but is a {\bf four-vector} field for the two alternative equations \cite{A39}, based on the {\it tensor representation of the Dirac field (TRD)} \cite{A37,A40};\\

\noindent and where \Couleur{$D_\mu $} is a covariant derivative, associated with a specific {\it connection:} one for each of the three equations.

}

\subsection{Dirac equation with vector wave function} 
{ 
For Dirac's original equation, the wave function is a (bi){\it spinor}.
This is due to the Dirac matrices \Couleur{$\gamma ^\mu $} being assumed Lorentz-invariant. However, a {\it matrix} usually does {\it not} remain invariant after a coordinate change.\\

\noindent In {\bf TRD}, the wave function $\psi $ is a (4-){\bf vector} instead, and the set of the four Dirac matrices $\gamma ^\mu$ builds a third-order {\bf tensor}: after a coordinate change,
\be \label{TRD-transform}
\Couleur{\psi'^\mu = \frac{\partial x'^\mu }{\partial x^\sigma  }\psi ^\sigma},\qquad 
\Couleur{\gamma'^{\mu \rho  }_\nu \equiv \left(\gamma'^\mu \right)^\rho _{\ \ \nu}= \frac{\partial x'^\mu }{\partial x^\sigma  }\frac{\partial x'^\rho  }{\partial x^\tau  }\frac{\partial x^\chi  }{\partial x'^\nu }\gamma ^{\sigma \tau }_\chi}.
\ee
This too leaves the {\it usual} Dirac equation covariant \cite{A37}. Moreover, the associated QM predictions in a flat spacetime are left {\bf unchanged}, for:\\

1) {\it The explicit (coordinate) expression of the Dirac equation is the same} as with the standard (spinor) transformation behaviour, and\\

2) {\it There is no influence of the possible set of Dirac matrices} \cite{A40}.
} 

 \subsection{The three different connections}

\vspace{2mm}

\begin {itemize}

\item For the two alternative equations (TRD), this is an affine connection:
\be\label{D_mu psi^nu}
\Couleur{(D_\mu \psi)^\nu \equiv \partial_\mu \psi^\nu  +\Delta ^\nu _{\rho \mu }\psi ^\rho}.\\
\ee

\begin {itemize}
\item For one of the two TRD equations (TRD-1), this is the Levi-Civita connection. I.e., $\Couleur{\ \Delta ^\nu _{\rho \mu }=\left\{^\nu _{\rho \mu } \right\} \ }$, the Christoffel symbols associated with the spacetime metric \Couleur{$g_{\mu \nu}$}. The corresponding gravitational Dirac equation {\it obeys the equivalence principle} \cite{A39}.\\
\item For TRD-2, the connection \Couleur{$\Delta $} is defined from the {\it spatial} Levi-Civita connection in an assumed {\it preferred reference frame} \cite{A39}.\\
\end {itemize}

\item For the standard equation (DFW), one uses the ``spin connection", which {\it depends on the \Couleur{$\gamma ^\mu$} matrices} and is generally complex \cite{BrillWheeler1957+Corr}.\\
\end {itemize}


\section{A common tool: the hermitizing matrices}
\vspace{2mm}

{\it For TRD}, the set \Couleur{$(\gamma  ^\mu )$} builds a tensor, hence {\it cannot be fixed.} This is true even for the ``flat" matrices \Couleur{$\gamma ^{\sharp \alpha}  $} if one defines 
\be\label{gamma-by-tetrad}
\Couleur{\gamma ^\mu = a^\mu_{\ \,\alpha}  \ \gamma ^{ \sharp \alpha}} 
\ee
with \Couleur{$a^\mu_{\ \,\alpha} $} an orthonormal tetrad. We must be able to use {\it any} possible set \Couleur{$(\gamma  ^\mu )$} of Dirac matrices. Note that, also for DFW, in which the gamma matrices are always defined through a tetrad by Eq. (\ref{gamma-by-tetrad}), one should study the influence of the choice of the ``flat" matrices \Couleur{$\gamma ^{\sharp \alpha}  $} and the tetrad $\Couleur{a^\mu_{\ \,\alpha}}$.\\

The solution is to use the {\it hermitizing matrix:} this is a \Couleur{$\ 4 \times 4$} matrix \Couleur{$A$} such that
\be\label{hermitizing-A}
\Couleur{A^\dagger = A, \qquad (A\gamma ^\mu )^\dagger = A\gamma ^\mu \quad \mu =0, ...,3},
\ee
where \Couleur{$M^\dagger\equiv M^{*\,T}$} = Hermitian conjugate of matrix \Couleur{$M$}. For the usual sets \Couleur{$(\gamma ^{\sharp \alpha})  $} (Dirac's, ``chiral", Majorana), \Couleur{$A= \gamma ^{\sharp 0}$}.\\

The matrix \Couleur{$A$}, introduced by Bargmann, was studied mainly by Pauli \cite{Pauli1933}. The existence of \Couleur{$A$} (and $\Couleur{B}$: for the $\Couleur{\alpha^\mu}$ matrices) has been proved by us for any set \Couleur{$(\gamma ^\mu)$} in a general metric \cite{A40}.\\

 
\section{Definition of the probability current}

In a flat spacetime, the current is unambiguously defined as
\be \label{J-mu-standard-matrix}
\Couleur{J^\mu =  \psi ^\dagger A\gamma ^\mu \psi}.
\ee
The definition (\ref{J-mu-standard-matrix}) is generally-covariant, the current being indeed a {\it four-vector,} for TRD and for DFW as well. Thus it holds true in a curved spacetime \Couleur{$\mathrm{V}$}. (Then \Couleur{$\gamma ^\mu$}, \Couleur{$A$} depend on  \Couleur{$X \in \mathrm{V}$}.)\\

The current (\ref{J-mu-standard-matrix}) is {\it independent of the choice of the Dirac matrices:} if one changes one set \Couleur{$(\gamma ^\mu)$} for another one \Couleur{$(\tilde {\gamma }^\mu)$}, the second set can be obtained from the first one by a point-dependent {\it similarity transformation:}
\be \label{similarity-gamma}
\Couleur{\exists S=S(X) \in {\sf GL}(4,{\sf C}):\qquad \tilde{\gamma} ^\mu(X) =  S\gamma ^\mu(X) S^{-1}, \quad \mu =0,...,3},
\ee
With the change \Couleur{$\tilde{\psi}=S\psi $}, this leaves the current unchanged \cite{A40,A42}. \\

 \section{Condition for current conservation}
This is specified by the following result \cite{A42}:
\paragraph{Theorem 1.}\label{Theorem1} {\it Consider the general Dirac equation in a curved spacetime (\ref{Dirac-general}), thus either DFW or any of the two TRD equations. In order that any \Couleur{$\psi $} solution of (\ref{Dirac-general}) satisfy the current conservation
\be\label{current-conservation}
\Couleur{D_\mu J^\mu=0},
\ee 
it is necessary and sufficient that} 
\be\label{D_mu(B_mu)=0}
\Couleur{D_\mu (A\gamma ^\mu)=0}.\\
\ee

\paragraph{Corollary 1.}\label{Corollary1} {\it For DFW theory, the hermitizing matrix field \Couleur{$A(X) $} can be imposed to be the} constant {\it matrix \Couleur{$A^\sharp $}, i.e., a hermitizing matrix for the ``flat" matrices \Couleur{$\gamma ^{\sharp \alpha}  $} of Eq. (\ref{gamma-by-tetrad}). Then the current conservation applies to any solution of the DFW equation.}\\

 \section{Admissible coefficient fields}

\hyperref[Theorem1]{Theorem 1} means that not all possible coefficient fields \Couleur{$(\gamma ^\mu,A)$} of the Dirac equation are physically admissible, but merely the ones which, in addition to the anticommutation relation (\ref{Clifford}), satisfy the field equation (\ref{D_mu(B_mu)=0}) ensuring current conservation. Such systems we call {\it ``admissible."} \\

Example: in a {\it flat} spacetime, relevant fields \Couleur{$\gamma ^\mu $} are ones which are {\it constant in Cartesian coordinates} (and hence also the field \Couleur{$A$}). Then the condition for current conservation (\ref{D_mu(B_mu)=0}) is satisfied.\\

If one selects the gamma field [satisfying (\ref{Clifford})] ``at random," the condition (\ref{D_mu(B_mu)=0}) and the current conservation do not generally apply to the solutions of the Dirac equation (\ref{Dirac-general}) {\it even in a flat spacetime}---{\it except for DFW.} \\

 \section{The Hamiltonian is frame-dependent}

The Dirac equation (\ref{Dirac-general}) can be put into Schr\"odinger form:
\be \label{Schrodinger-general}
\Couleur{i \frac{\partial \psi }{\partial t}= \mathrm{H}\psi,\qquad (t\equiv x^0)},
\ee
with
\be \label{Hamilton-Dirac-general}
\Couleur{\mathrm{H} \equiv  m\alpha  ^0 -i\alpha ^j D _j -i(D_0-\partial_0)}, 
\ee
where
\be \label{alpha}
\Couleur{\alpha ^0 \equiv \gamma ^0/g^{00}}, \qquad \Couleur{\alpha ^j \equiv \gamma ^0\gamma ^j/g^{00}}.
\ee
In order that the Hamiltonians $\Couleur{\mathrm{H}}$ and $\Couleur{\mathrm{H}'}$ before and after a coordinate change be equivalent operators, the coordinate change must be a {\bf spatial} change:
\be
 \Couleur{x'^0=x^0,\ x'^j=f^j((x^k))}. 
\ee
Then, both sides of the Schr\"odinger equation (\ref{Schrodinger-general}) behave as a scalar for DFW, and as a vector for TRD. Thus {\bf \Couleur{H} depends on the reference frame}, i.e., on the three-dimensional congruence of world lines (observers) which is considered. This frame dependence of the Hamiltonian and the resulting necessity of restricting oneself to spatial coordinate changes is a general result that holds true for other wave equations. In Ref. \cite{A42}, we give a precise definition of a {\it reference frame} in this context, and we study it mathematically.\\

 \section{Hermiticity condition for the Hamiltonian}
%
The Hilbert scalar product is fixed by the following result \cite{A42}:
\paragraph{Theorem 5.}\label{Theorem5} {\it A} necessary {\it condition for the scalar product of time-independent wave functions to be time independent and for the Hamiltonian \Couleur{$\mathrm{H}$} to be a Hermitian operator, is that the scalar product should be}
\be \label{Hermitian-sigma=1-g}
\Couleur{(\psi  \mid \varphi  ) \equiv \int_{{\sf R}^3} \psi^\dagger A\gamma ^0  \varphi \ \ \sqrt{-g}\ \dd^ 3{\bf x}}.
\ee

\vspace{3mm}
The hermiticity condition for the Hamiltonian, w.r.t. this scalar product, is then given by \cite{A42}:

\paragraph{Theorem 6.}\label{Theorem6} {\it Assume that the coefficient fields \Couleur{$(\gamma ^\mu ,A)$} satisfy the two admissibility conditions (\ref{Clifford}) and (\ref{D_mu(B_mu)=0}). In order that the Dirac Hamiltonian (\ref{Hamilton-Dirac-general}) be Hermitian for the scalar product (\ref{Hermitian-sigma=1-g}), it is necessary and sufficient that}
\be\label{hermiticity-condition-}
\Couleur{\partial _0 \left(\sqrt{-g}\,  A \gamma^0 \right)=0}.  
\ee
\vspace{4mm}
 \section{For DFW, hermiticity is not stable under a local similarity transformation}
For DFW, {\it all} local similarity transformations are admissible, since condition (\ref{D_mu(B_mu)=0}) is always satisfied [with the choice $\Couleur{A(X)\equiv A^\sharp}$: see \hyperref[Corollary1]{Corollary 1}]. In contrast, for TRD, condition (\ref{D_mu(B_mu)=0}) is quite demanding.\\

For DFW, in very general coordinates, the tetrad $\Couleur{(a^\mu _{\ \,\alpha} )}$ may be chosen to satisfy \Couleur{$a^0_{\ \,j}=$}\Couleur{$\,0$}. Taking for ``flat" matrices the standard Dirac matrices \Couleur{$\gamma ^{\sharp \, \alpha  }$}, the hermiticity condition (\ref{hermiticity-condition-}) then reduces to Leclerc's \cite{Leclerc2006} :
\be\label{Leclerc-2}
\Couleur{\partial _0 (\sqrt{-g \,g^{00}})=0}.
\ee
But, after a local similarity \Couleur{$S$}, the condition (\ref{hermiticity-condition-}) becomes
\be \label{hermiticity-similarity}
\Couleur{\partial_0(\sqrt{-g\,g^{00}} \,S^\dagger  S)=0 },
\ee
which obviously {\it cannot} be satisfied if (\ref{Leclerc-2}) is, and if moreover \Couleur{$S^\dagger S=F(t)$}.\\

 \section{Conclusion}
\vspace{2mm}
{
\begin{itemize}
\item {\it Two} new gravitational Dirac equations were previously derived from wave mechanics \cite{A39}. One obeys the equivalence principle, the other one has a preferred reference frame. Both see the wave function as a {\it vector}. This leaves QM associated with the Dirac equation of special relativity unchanged, i.e., for Minkowski spacetime in Cartesian coordinates \cite{A40}.\\

\item The three gravitational Dirac equations have been studied in a common framework, using the hermitizing matrix (field) $\Couleur{A}$ \cite{A42}:

\begin{itemize}

\item The current conservation asks for the matrix equation $\Couleur{D_\mu (A\gamma ^\mu)=0}$. Thus, not all coefficient fields of the Dirac equation are admissible.\\

\item The hermiticity condition is $\Couleur{\partial _0 \left(\sqrt{-g}\,  A \gamma^0 \right)=0}$. This is frame dependent, as is natural indeed. We gave a formal definition of a reference frame. \\

\item For DFW, this condition is not stable under admissible similarity transformations: the standard version of the gravitational Dirac equation seems to have a uniqueness problem. We are currently pursuing the study to clarify this.
\end{itemize}

\end{itemize}

}

\end{document}